\documentclass[lettersize,journal]{IEEEtran}
\usepackage{amsmath,amssymb,amsfonts}
\usepackage{algorithmic}
\usepackage{algorithm}
\usepackage{array}
\usepackage[caption=false,font=normalsize,labelfont=sf,textfont=sf]{subfig}
\usepackage{textcomp}
\usepackage{stfloats}
\usepackage{url}
\usepackage{verbatim}
\usepackage{graphicx}
\usepackage{cite}

\newtheorem{thm}{Theorem}

\newtheorem{rem}{Remark}
\newtheorem{assu}{Assumption}
\newtheorem{prob}{Problem Statement}

\begin{document}

\title{Learning swarm behaviour from a flock of homing pigeons using inverse optimal control}

\author{Afreen Islam
        % <-this % stops a space
\thanks{Afreen Islam is with the Center for Space Research, and the Department of Mechanical and Materials Engineering, University College Dublin, Belfield, Dublin D04 C1P1, Ireland. \textit{Corresponding author: Afreen Islam}, (e-mail: afreenislam@gmail.com)}% <-this % stops a space
%\thanks{Manuscript received April 19, 2021; revised August 16, 2021.}}
}

% The paper headers
%\markboth{Journal of \LaTeX\ Class Files,~Vol.~14, No.~8, August~2021}%
%{Shell \MakeLowercase{\textit{et al.}}: A Sample Article Using IEEEtran.cls for IEEE Journals}

%\IEEEpubid{0000--0000/00\$00.00~\copyright~2021 IEEE}
% Remember, if you use this you must call \IEEEpubidadjcol in the second
% column for its text to clear the IEEEpubid mark.

\maketitle

\begin{abstract}
In this work, Global Position System (GPS) data from a flock of homing pigeons are analysed. The flocking behaviour of the considered homing pigeons is formulated as a swarm optimal trajectory tracking control problem. The swarm problem in this work is modeled with the idea that one or two pigeons at the forefront lead the flock. Each follower pigeon is assumed to follow a leader pigeon immediately ahead of themselves, instead of directly following the leaders at the forefront of the flock. The trajectory of each follower pigeon is assumed to be a solution of an optimal trajectory tracking control problem. An optimal control problem framework is created for each follower pigeon. An important aspect of an optimal control problem is the cost function. A minimum principle based method for multiple flight data is proposed, which can help in learning the unknown weights of the cost function of the optimal trajectory tracking control problem for each follower pigeon, from flight trajectories' information obtained from GPS data.
\end{abstract}


\section{Introduction}
Swarming behaviour in animals is one of the most fascinating phenomena in nature where each individual animal interacts with their close neighbours to produce large-scale complex behaviour \cite{satz2020rules}. It  has been observed in trilobites dating back about 480 million years and has been perfected and optimised over millions of years of evolution. Today swarming behaviour can be observed in flocks of migratory birds, insects, fish, and quadrupedal mammals. Humans have since a long time tried to understand swarming behviour from nature. Beer \cite{beer1958composition} questioned whether a group of birds have any distinctive utility or whether they were merely haphazard organizations.

Swarming behaviour offers individual members defensive advantages and more efficient navigation when compared to solitary behaviour \cite{vine1971risk}. Vine et al \cite{vine1971risk} found that a circular grouping provided the best predator avoidance strategy against visual predators. Weimerskirch et al \cite{weimerskirch2001energy} found that the heart rates of birds in line formations were $11-15\%$ lower than that of a bird flying alone. Lissaman and Schollegenberger \cite{lissaman1970formation} used aerodynamic theory to find that a group of 25 birds (species not mentioned) flying in a V formation would have $71\%$ more range than a single bird. In the work by Papadopoulou et al, GPS trajectories of flocks of flying pigeons attacked by a robotic falcon were analysed \cite{papadopoulou2022self}.  The results demonstrated how during the self-organization process, flock members increase their consensus over which direction to escape and turn collectively as the predator approaches close to the flock. The analysis indicate that coordination among flock members, combined with simple escape rules, reduces the cognitive costs of tracking the predator while flocking.

The understanding of swarming behaviour in recent times has brought together experts from different scientific fields such as biology, engineering, mathematics and statistics.  The work of Dieck et al \cite{dieck2012dynamical} was among the first to use computational modeling methods to infer flocking behaviour from experimental data. Their work demonstrates the existence of a basic distance dependent attraction/repulsion relationship between members of a flock of pigeons. They showed that such trivial rules are sufficient to explain collective behaviour observed in nature. Emlen \cite{emlen1952flocking} suggested that both flocking and structure of a flock were the result of an interplay of attractive and repulsive behavioural forces. Reynolds also suggested that coordinated flocking is the result of extremely simple behavioural rules followed by each bird in the group \cite{reynolds1987flocks}. 

Line formations are seen more commonly in large birds that fly in regular lines or queues, such as geese, cormorants or duck, whereas Cluster formations, which have a three dimensional structure like a sphere, are observed in smaller birds such as pigeons, starlings and smaller shorebirds \cite{heppner1974avian}, \cite{bajec2009organized}. 

There have been studies where flocks of pigeons were considered.  Nichols \cite{nichols1931notes} proposed that a change in direction of a flock of flying pigeons was related to a change in positional leadership. In the work by Nagy et al, GPS data of a flock of homing pigeons were analysed to understand the hierarchy of individual members of the flock \cite{nagy2010hierarchical}. It was suggested that certain birds have a natural tendency to act as leaders irrespective of navigational context. It was established that in general, individuals at the front of the flock tend to assume leadership roles. Yomosa et al found that an important aspect of collective motion is group decision making \cite{yomosa2015coordinated}. It was found that there is a tendency to maintain longitudinal relative position when the flock turns \cite{yomosa2015coordinated}. During homing flights, flock members are forced to resolve potentially conflicting navigational preferences while flying as a flock \cite{flack2013robustness}. An important finding of the work by Flack et al is that training did not consistently result in an increase in birds’ leadership ranks and that in general, the nature of leader-follower interactions between trained and untrained birds remain unaffected \cite{flack2013robustness}. 

In recent years, there has been a tremendous development in understanding of swarm behaviour in a flock of birds.  Despite the progress made, there are open questions about the philosophy behind how the birds generate their trajectories while flying as a flock. It has been hypothesised that natural movements and locomotion are a result of an optimal control problem \cite{flash1985coordination}, \cite{arechavaleta2008optimality}. The Inverse Optimal Control (IOC) problem \cite{mombaur2010human} \cite{johnson2013inverse} from the control systems community and the Inverse Reinforcement Learning (IRL) problem \cite{ng2000algorithms}, \cite{abbeel2004apprenticeship} from the Machine Learning community both identify the cost/ reward function as the most succinct, robust and transferable definition of a task \cite{ng2000algorithms}. Mombaur et al formulated the human walking problem as an optimal control problem \cite{mombaur2010human} and attempted to transfer biological locomotion from humans to humanoid robots using a bilevel IOC method\cite{mombaur2010human}.  Westermann et al analysed the human jumping movements and formulated the problem as an inverse optimal control problem \cite{westermann2020inverse}. Jin et al formulated the human squatting task as an optimal control problem with multi-phase cost functions and hence, put forward a minimum-principle based IOC method to recover the cost functions for each phase of the squatting task. Pinsler et al formulates a flock of birds as a maximum entropy inverse reinforcement learning problem \cite{pinsler2018inverse}.

There are numerous algorithms for the control of multi-agent systems. Robust adaptive control algorithms for a swarm of multi-UAV systems have been proposed by Su et al \cite{su2023robust}, Nguyen et al\cite{nguyen2019robust}, Shu et al \cite{Shu2021robust}, Wang et al \cite{Wang2023robust}. Shen and Wei proposes a leader-follower algorithm for multi-UAV flocking control \cite{shen2022multi}. Wu et al proposes a semi-definite programming based robust control algorithm for containment coordination in a swarm system\cite{wu2021sdp}. Considering the importance of decision making in multi-agent systems, Rizk et al surveys existing works on control of multi-agent systems \cite{rizk2018decision}. Na et al puts forward a deep reinforcement learning bio-inspired algorithm for collision avoidance in a swarm system \cite{na2022bioinspired}. Sliding mode control for multi-agent systems have been proposed by Li et al\cite{li2022event}, Li et al\cite{li2022sliding} and Dong et al\cite{dong2016sliding}. Optimal control algorithms have also been put forward for multi agent-systems, for example, by Wang et al \cite{wang2012integrated} and Ertug et al \cite{ertug2019optimal}. 

However, the motion of a swarm of birds has not yet been formulated as a trajectory tracking optimal control problem. IOC can be seen as an approach for biomimicry \cite{islam2025uniqueness}. In this work, the swarm behaviour of birds is formulated as an optimal trajectory tracking control problem. To faciliate learning from a swarm of homing pigeons, the trajectories are analysed using inverse optimal control techniques to learn the weights of the unknown cost function of the optimal control problem. Learning the unknown weights of the optimal trajectory tracking control problem from a flock of pigeons can lay the foundation for replicating the swarming behaviour of birds from nature into a swarm of drones or other Unmanned Aerial Vehicle (UAV) system.

The main contributions of this work are:
\begin{enumerate}
    \item The swarming behaviour of a flock of homing pigeons is formulated as a trajectory tracking optimal control problem.
    \item The trajectories of the flock of pigeons are analysed and an IOC problem is formulated to learn the unknown weights of the cost function of the optimal control problem of the follower pigeons in the flock.
    \item The hard-constrained minimum principle based IOC method is formulated to understand the swarm behaviour and extended for multiple flight trajectories for the follower pigeons in the flock.
    \item This work lays the foundation for learning the optimal flocking behaviour from a swarm of birds, which can be then be applied in swarm algorithms for multi UAV systems.
\end{enumerate}

\section*{Kinematics model of pigeon flight}
\begin{figure}
    \centering
\includegraphics[scale=0.5]{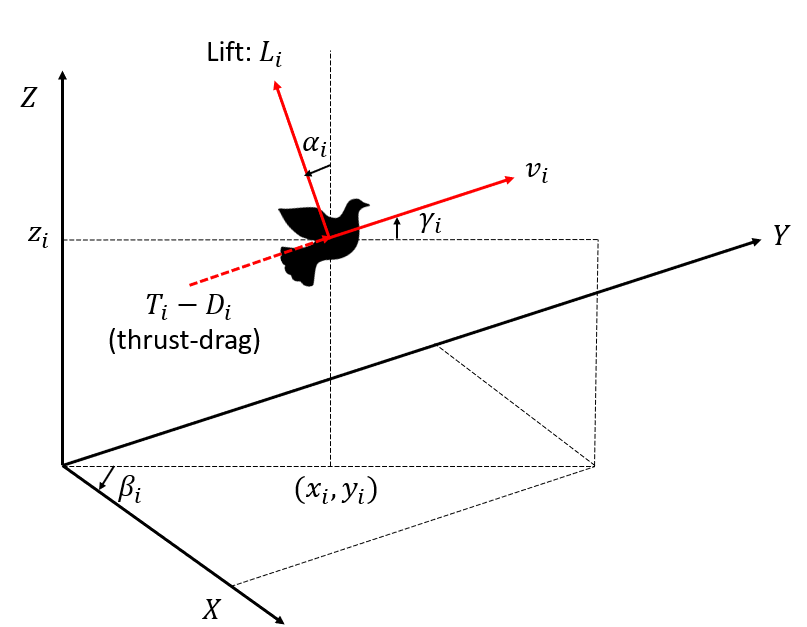}
    \caption{Kinematics diagram for the flight of a pigeon}
    \label{fig:kinematics}
    \end{figure}

For simplicity and brevity, we model the kinematics of a pigeon flight in a similar way to an unmanned aerial vehicle (UAV) as in Shen et al \cite{shen2022multi}. Let the motion of the bird be as shown in Fig. \ref{fig:kinematics}. $T_i$, $D_i$ and $L_i$ are respectively the thrust, drag and the lift on the considered $i^{\hbox{th}}$ pigeon in a flock of multiple pigeons. $v_i$ is the ground velocity of the $i^{\hbox{th}}$ pigeon. $\alpha_i$, $\beta_i$ and $\gamma_i$ are respectively the banking, the heading and the flight path angles of the $i^{\hbox{th}}$ pigeon. Similar to the model in Shen et al \cite{shen2022multi}, we let $\eta_x=\frac{T_i-Di}{mg}$ and $\eta_f=\frac{L_i}{mg}$, where $m$ is the mass of the pigeon and $g$ is the acceleration due to gravity. In these equations, $x_i$, $y_i$ and $z_i$ are the positions of the $i^{\hbox{th}}$pigeon in x, y and z axis, respectively, while $v_i$ indicates the resultant velocity vector. The kinematics model\cite{shen2022multi} of the considered pigeon is:
\begin{equation}
  \begin{cases}
 \dot{x}_i= v_i \hbox{cos}\gamma_i \hbox{cos}\beta_i,\\
 \dot{y}_i= v_i \hbox{cos}\gamma_i \hbox{sin}\beta_i,\\
 \dot{z}_i= v_i \hbox{sin}\gamma_i,\\
 \dot{v}_i=g(\eta_x-\hbox{sin}\gamma_i),\\
 \dot{\gamma}_i= \frac{g}{v_i}(\eta_f\hbox{cos}\alpha_i-\hbox{cos}\gamma_i),\\
 \dot{\beta}_i= \frac{g}{v_i\hbox{cos}\gamma_i}\eta_f\hbox{sin}\alpha_i,
  \end{cases}
\end{equation}

For the sake of simplicity, we use a linearised double integrator model \cite{shen2022multi}, \cite{su2023robust}.

\begin{equation}
  \begin{cases}
 \ddot{x}_i= a_{x_i} ,\\
 \ddot{y}_i= a_{y_i},\\
 \ddot{z}_i= a_{z_i},\\
  \end{cases}
  \label{double_integrator}
\end{equation}
where $a_{xi}$, $a_{yi}$ and $a_{zi}$ are the acceleration along x, y and z directions of the $i^{\hbox{th}}$ pigeon, whose kinematics is considered in Fig. \ref{fig:kinematics}.

\section*{Formulation of a flock of pigeons as a swarm optimal control problem}
\begin{figure}
    \centering
\includegraphics[scale=0.5]{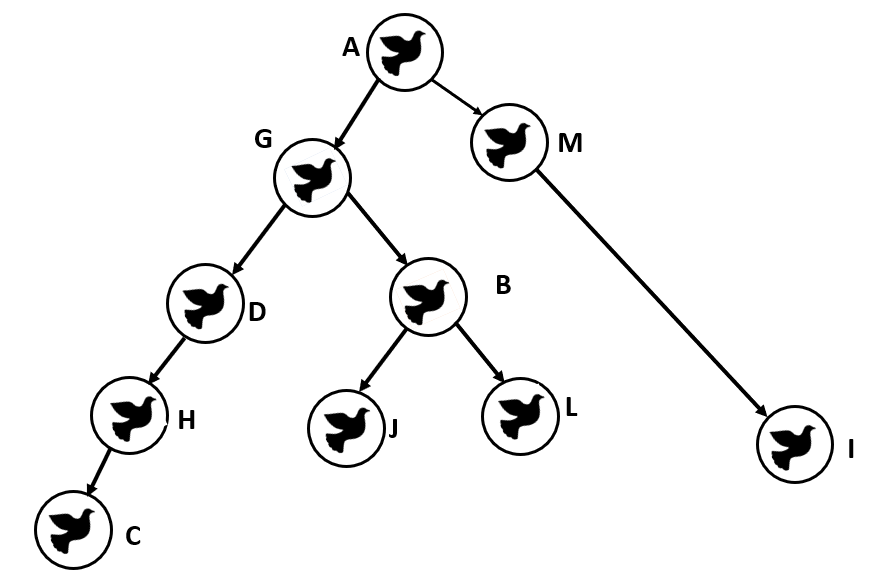}
    \caption{Hierarchy of the pigeons in the flock}
    \label{fig:hierarchy}
    \end{figure}

\begin{figure}
    \centering
\includegraphics[scale=0.5]{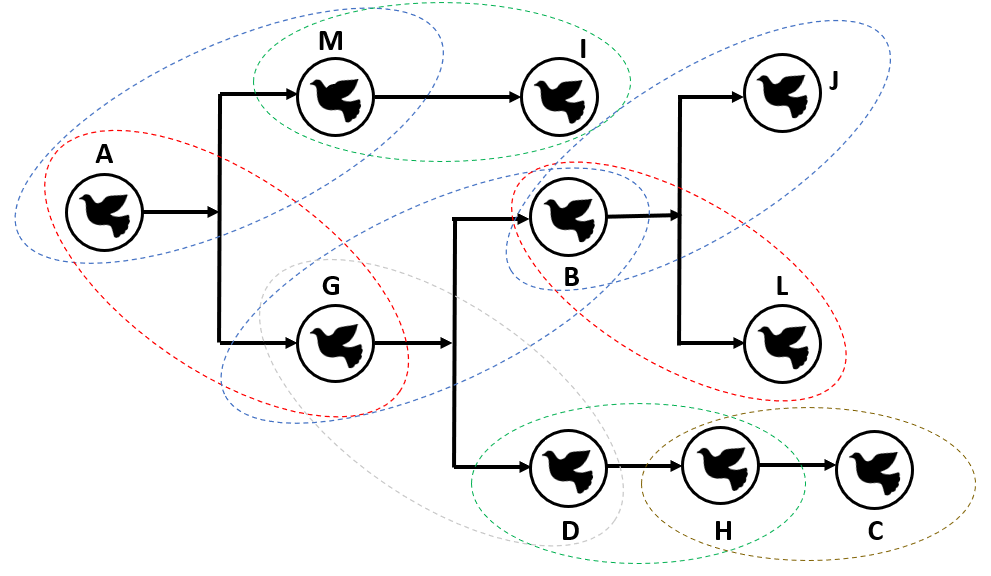}
    \caption{Leader-follower pairs in the flock of pigeons indicated by dotted circular contours}
    \label{fig:leader-follower}
    \end{figure}

In the seminal paper of Nagy et al \cite{nagy2010hierarchical}, the leader-follower behaviour has been establised for a flock of homing pigeons. In these flocks of pigeons carrying out free-flights or homing flights, one or a very few pigeons act as leaders and rest of the pigeons follow the leader(s). In the book by Satz\cite{satz2020rules}, it was explained how instead of following the leaders at the forefront, each bird in a swarm follows a bird immediately ahead of themselves. These important concepts will be used for formulating a swarm optimal control problem for a flock of homing pigeons. For free-flight data of homing pigeons from Nagy et al\cite{nagy2010hierarchical}, 10 pigeons are considered, and the hierarchy will be considered according to Figure 2b in Nagy et al\cite{nagy2010hierarchical}. Let us consider Fig. \ref{fig:hierarchy}, to explain the swarm optimal control problem. To understand the swarm behaviour of the pigeon flocks in Fig. \ref{fig:hierarchy}, we consider that pigeon A at the forefront is the leader of the entire flock. We then consider that pigeons G and M follow pigeon A. Thereafter, it is considered that Pigeons B and D follow pigeon G. Pigeon I follows pigeon M. In the next hierarchy, it is assumed that pigeon H follows pigeon D, while pigeon C follows pigeon H. It is next assumed that pigeons J and L follow pigeon B. The leader-follower pairs in the flock of pigeons considered are also shown in Fig. \ref{fig:leader-follower} and Table \ref{leader_follower_pair}. In Fig. \ref{fig:leader-follower}, each leader-follower pair is shown in a dotted circular contour.

    \begin{table}
   \begin{center} 
    \begin{tabular}{ |c|c|c|c| }
 \hline
\textbf{Sl No}  &  \textbf{Leader} & \textbf{Follower} & \textbf{Time Delay} \\
 \hline
 1  & A  &  M & 0.2 s \\
   \hline
  2 & A & G & 0.6 s\\
  \hline
    3 & G & B & 0.2 s\\
    \hline
    4 & G & D & 0.2 s\\
    \hline
    5 & M & I & 0.6 s\\
    \hline
    6 & B & J & 0.2 s\\
    \hline
    7 & B & L & 0.2 s \\
    \hline
    8 & D & H & 0.2 s\\
    \hline
     9 & H & C & 0.2 s\\
 \hline

 \end{tabular}

 \caption{\label{leader_follower_pair} Leader-follower pairs for the flock of pigeons}

\end{center}
\end{table}

For each follower pigeon, the desired trajectory is the trajectory of the corresponding leader pigeon, delayed in time. For instance, for the leader-follower pigeon pair A-M, it is considered that pigeon M solves an optimal trajectory tracking problem. The desired trajectory of M is the trajectory of A delayed by 0.2 s. This choice is based on Figure 2b of Nagy et al \cite{nagy2010hierarchical}, where the delay time of motion between pigeons A and M is indicated as 0.2 s. The optimal trajectory tracking problem for each follower pigeon will now be formulated and for that we require a mathematical model of each follower pigeon.

\subsection*{State-space model for each follower pigeon in a flock}
For the sake of simplicity, we will consider the double integrator model in (\ref{double_integrator}), similar to Su et al \cite{su2023robust}. From the double integrator model (\ref{double_integrator}) of a pigeon flight, we deduce it's state-space model. The state-space model of each follower pigeon $i\in\{M,G,B,D,H,L,I,C,J\}$ is given as:
\begin{equation}
    \dot{X_i}(t)=\mathcal{A}X_i(t)+\mathcal{B}U_i(t),
\end{equation}
where $X_i(t)=\begin{bmatrix}
  x_i(t) & y_i(t) & z_i(t) & v_{x_i}(t)& v_{y_i}(t) & v_{z_i}(t)
\end{bmatrix}^T$ is the state of the $i^{\hbox{th}}$ follower pigeon,\\ $U_i(t)=\begin{bmatrix}
    u_{x_i}(t) & u_{y_i}(t) & u_{z_i}(t)
\end{bmatrix}^T=\begin{bmatrix}
    a_{x_i}(t) & a_{y_i}(t) & a_{z_i}(t)
\end{bmatrix}^T$ is the control (acceleration) of the $i^{\hbox{th}}$ follower pigeon, $\mathcal{A}=\begin{bmatrix}
    \mathbf{0}_{3\times 3} & \mathbf{I}_{3\times 3}\\
    \mathbf{0}_{3\times 3} & \mathbf{0}_{3\times 3}
\end{bmatrix}$ and $\mathcal{B}=\begin{bmatrix}
    \mathbf{0}_{3\times 3} & \mathbf{I}_{3\times 3}
\end{bmatrix}^T$. Here, $x_i(t)$, $y_i(t)$, $z_i(t)$ denote the x, y and z coordinates respectively of the $i^{\hbox{th}}$ follower pigeon at time $t$. $v_{x_i}(t)$, $v_{y_i}(t)$, $v_{z_i}(t)$ denote the velocity of the $i^{\hbox{th}}$ pigeon at time $t$ along x, y and z directions, respectively. $a_{x_i}(t)$, $a_{y_i}(t)$, $a_{z_i}(t)$ denote the acceleration of the $i^{\hbox{th}}$ pigeon at time $t$ along x, y and z directions, respectively.  
\subsection*{Optimal trajectory tracking control problem for each follower pigeon in the flock}
It will now be assumed that the trajectories of each follower pigeon is generated by solving an optimal trajectory tracking problem. It has been hypothesised that natural movements and locomation are optimal \cite{flash1985coordination}, \cite{arechavaleta2008optimality}. Hence, formulating each leader-follower pigeon pair in a flock as an optimal control problem is a logical choice. The optimal trajectory tracking control problem for the 
$i^{\hbox{th}}$ follower pigeon is given as
\begin{equation}
\begin{aligned}
& \underset{U_i(t)}{\text{minimize}}
& & (X_{T_i}(t)-X_i(t))^TQ_i(X_{T_i}(t)-X_i(t))\\ & & &+U_i(t)^TR_iU_i(t)\\
& \text{subject to}
& & \dot{X_i}(t)=\mathcal{A}X_i(t)+\mathcal{B}U_i(t).
\label{original_optimal_problem}
\end{aligned}
\end{equation}
 
\begin{figure*}[t]
\begin{equation}
X_{T_i}(t)=\begin{bmatrix}
  x_{T_i}(t) & y_{T_i}(t) & z_{T_i}(t) & v_{x_{T_i}}(t)
 & v_{y_{T_i}}(t) & v_{z_{T_i}}(t)
\end{bmatrix}^T
\label{desired_trajectory}
\end{equation}
\hrulefill
\end{figure*}
In the optimal control problem (\ref{original_optimal_problem}), $X_{T_i}$ shown in equation (\ref{desired_trajectory}) at the top of pg. \pageref{desired_trajectory} is the desired trajectory of the $i^{\hbox{th}}$ follower pigeon, where $x_{T_i}(t)$, $y_{T_i}(t)$ and $z_{T_i}(t)$ are the desired x, y and z positions and $v_{x_{T_i}}(t)$, $v_{y_{T_i}}(t)$ and $v_{z_{T_i}}(t)$ are the desired velocities along the x, y and z directions. $Q_i\geq0$ and $R_i>0$ are the respective weighing matrices on the tracking error $(X_{T_i}(t)-X_i(t))$ and the control efforts $U_i(t)$. $X_{T_i}(t)$ is chosen as the time-delayed trajectory of the corresponding leader in the leader-follower pair under consideration. For example, for the pair A-M in Fig. \ref{fig:hierarchy}, the leader pigeon is A and the follower pigeon is M. The desired trajectory of M, $X_{T_M}(t)$, is the time-delayed trajectory of pigeon A with the delay time as 0.2 s as shown in equation (\ref{XTmt}) as shown at the top of pg. \pageref{XTmt}.
\begin{figure*}[t!]
\begin{equation}
X_{T_M}(t)=\begin{bmatrix}
  x_{A}(t-0.2) & y_{A}(t-0.2) & z_{A}(t-0.2) & v_{x_A}(t-0.2) & v_{y_A}(t-0.2) & v_{z_A}(t-0.2)
\end{bmatrix}^T
\label{XTmt}
\end{equation}
\hrulefill
\end{figure*}

\begin{assu}
    It is assumed that the weighing matrices $Q_i\geq 0$ and $R_i>0$ are diagonal.
    \label{diagonal_wts}
\end{assu}

\section*{Background: Inverse optimal control problem}
The inverse optimal control problem is an approach of recovering the unknown cost function of an existing optimal control problem by measuring its state and control trajectories. Natural movements and locomotion have been hypothesized to be optimal \cite{flash1985coordination}, \cite{arechavaleta2008optimality}. Since the cost function has been identified to be the most succinct, robust and transferrable definition of a task \cite{ng2000algorithms}, the IOC problem can be seen as an approach for biomimicry \cite{islam2025uniqueness}.

For a general nonlinear optimal control problem
\begin{equation}
\begin{aligned}
& \underset{u}{\text{minimize}}
& &  \int_{t_0}^{t_f} \Phi(x(t),u(t)) dt, \\
& \text{subject to}
& & \dot{x}(t)=f(x(t),u(t)),\\
& & & x(t_0)=x_0,
\label{optimal_problem}
\end{aligned}
\end{equation}
$\Phi(x(t),u(t))$ is the cost function and $f(x(t),u(t))$ represents the system dynamics of a general nonlinear system. 

\begin{assu}
It is assumed that $f \in C^1$, that is, continuously differentiable \cite{johnson2013inverse}.
\label{assu_diff}
\end{assu}
The above assumption results in the system, $\dot{x}(t) = f(x(t), u(t))$  in problem (\ref{optimal_problem}) to be well-posed, that is, for every initial condition, $x_0$,
and every admissible control, $u(t)$, the equation $\dot{x}(t) = f(x(t), u(t))$, in problem (\ref{general_optimal_problem}) has a unique solution, $x(t)$, with $t \in [t_0, t_f ]$ \cite{johnson2013inverse}. 
\begin{assu}
It is assumed that the system in problem (\ref{optimal_problem}) is controllable \cite{islam2025uniqueness}.
\label{controllable}
\end{assu}

\begin{assu}
    It is assumed that $\Phi(x(t),u)(t))$ is positive semi-definite in $x(t)$ and $u(t)$, for all $t\ge 0$ \cite{islam2025uniqueness}.
    \label{assu_pd}
\end{assu}

%\begin{assu}
%It is assumed that the basis function, $\phi(t,x(t),u(t))$, is known, while the weights, $c$, of the cost function are unknown. However, it is assumed that the last element, $c_k$, of the vector, $c$, is known. 
%\end{assu}

\begin{assu}
    It is assumed that problem (\ref{optimal_problem})  is a non-singular problem \cite{islam2025uniqueness}.
    \label{assu_nonsingular}
\end{assu}

At this stage, we will define the IOC problem statement.
\begin{prob}
    Given the system dynamics $\dot{x}(t)=f(x(t),u(t))$, and a specific system trajectory $x(t)$ and control signal $u(t)$, $\forall t_0\le t\le t_f$, measured from the primary optimal control problem (\ref{general_optimal_problem}), the objective of the IOC problem is to find the unknown cost function, $\Phi(x(t),u(t))$.
    \label{goal_ioc}
\end{prob}

We next consider an important assumption.
\begin{assu}
     The cost function $\Phi(x(t),u(t))$ can be expressed as a weighted sum of known basis functions as \cite{mombaur2010human}:
     \begin{equation}
        \Phi(x(t)u(t)=\sum_{k=0}^r c_k\phi_k(x(t),u(t)).
     \end{equation} 
    Thus, if the weights $c_k$'s and the basis functions $\phi_k$'s are written in a vector form, $\Phi(x(t)u(t))$ can be expressed as
    \begin{equation}
     \Phi(x(t),u(t))=c^T\phi(x(t),u(t)),   
    \end{equation}
    where $c=\begin{bmatrix}
         c_1 & c_2 & \ldots & c_r
     \end{bmatrix}^T$ and $\phi(x(t),u(t))=\begin{bmatrix}
         \phi_1(x(t),u(t)) & \phi_2(x(t),u(t)) & \ldots & \phi_r(x(t),u(t))
     \end{bmatrix}^T$.
     \label{weighted}
\end{assu}
Using Assumption \ref{weighted}, the optimal control problem (\ref{optimal_problem}) becomes
 \begin{equation}
\begin{aligned}
& \underset{u}{\text{minimize}}
& &  \int_{t_0}^{t_f} c^T\phi(x(t),u(t)) dt, \\
& \text{subject to}
& & \dot{x}(t)=f(x(t),u(t)),\\
& & & x(t_0)=x_0,
\label{general_optimal_problem}
\end{aligned}
\end{equation}
Hence, the goal of the IOC problem now changes to finding the unknown weight vector $c$ instead of the cost function $\Phi(x(t),u(t))$. The IOC problem statement is now refined in the following way \cite{islam2025uniqueness}.
\begin{prob}
    Given the system dynamics $\dot{x}(t)=f(x(t),u(t))$, basis function vector $\phi(x(t),u(t) $and a specific system trajectory $x(t)$ and control signal $u(t)$, $\forall t_0\le t\le t_f$, measured from the primary optimal control problem (\ref{general_optimal_problem}), the objective of the IOC problem is to find the weight vector $c$.
    \label{goal_ioc_new}
\end{prob}
\subsection{The hard-constrained minimum principle-based IOC method}
Due to the intuitive nature of minimum-principle based methods, we will use the hard-constrained minimum principle based method \cite{molloy2019inverse} in this work. The minimum principle methods solve the IOC problem by assuming that the considered system is only approximately optimal and not strictly optimal. Optimality, here, is defined based on the necessary conditions of optimality or the Karush-Kuhn-Tucker conditions. In the next two sections,  this method will be extended for the swarm inverse optimal control problem. The hard-constrained method is a minimum principle based IOC approach \cite{molloy2019inverse}. The Hamiltonian function \cite{naidu2018optimal} for the problem (\ref{original_optimal_problem}) is:
\begin{align}
H(x(t),u(t),p(t))&=c^T\phi(x(t),u(t))+p^T(t)f(x(t),u(t)),
\label{hamiltonian}
\end{align}
where $p(t)$ is the costate trajectory. If $(x(t),u(t))$ are optimal, that is, $(x(t),u(t))=(x^*(t),u^*(t))$, where $(x^*(t),u^*(t))$ is the optimal solution of problem (\ref{general_optimal_problem}) for a given $c$, then there exists a costate trajectory, $p(t):\mathbb{R}\rightarrow\mathbb{R}^n$ (here, $p(t)=p^*(t)$, where $p^*(t)$ is the optimal costate trajectory) and it is necessary that \cite{johnson2013inverse}, \cite{athans1966optimal}:
\begin{align}
\dot{p}^T(t)+\nabla_x H(x(t),u(t),p(t))=\mathbf{0}, \label{cond1a}\\
\nabla_u H(x(t),u(t),p(t))=\mathbf{0}. \label{cond2a}
\end{align}
The following condition together with equations (\ref{cond1a}) and (\ref{cond2a}) is sufficient and necessary for optimality \cite{johnson2013inverse}:
\begin{equation}
\frac{\partial^2{H(t,x(t),u(t),p(t))}}{\partial{u^2}} > \mathbf{0}.
\label{suff_cond}
\end{equation}

For  the optimal control problem (\ref{general_optimal_problem}), the necessary conditions of optimality, equations (\ref{cond1a}) and (\ref{cond2a}) can be written as \cite{johnson2013inverse}, \cite{athans1966optimal}
\begin{align}
\dot{p}^T(t)+c^T\nabla_x\phi(x(t),u(t))+p^T(t)\nabla_xf(x(t),u(t))=\mathbf{0}, \label{cond1b}\\
c^T\nabla_u\phi(x(t),u(t))+p^T(t)\nabla_uf(x(t),u(t))=\mathbf{0}. \label{cond2b}
\end{align}

In the minimum principle based methods, it is assumed that the measurements, $(x(t),u(t))$, are only approximately optimal and not strictly optimal \cite{johnson2013inverse}. In a strictly optimal system, equations (\ref{cond1b}) and (\ref{cond2b}) hold strictly  whereas in an approximately optimal system, either or both of equations (\ref{cond1b}) and (\ref{cond2b}) hold only approximately \cite{johnson2013inverse}  \cite{molloy2019inverse}. 
In the hard-constrained method \cite{molloy2019inverse}, equation (\ref{cond1b}) is considered to be strictly satisfied while equation (\ref{cond2b}) is considered to be satisfied only approximately. In this method, the left-hand side of equation (\ref{cond2b}) is minimized assuming equation (\ref{cond1}) holds strictly. In the hard-constrained method \cite{molloy2019inverse}, constraints are imposed on the control trajectories in the primary problem. We will not consider constraints on the control trajectories in this work. The IOC problem for the method is:
\begin{equation}
\begin{aligned}
& \underset{c}{\text{minimize}}
& &   \int_{t_0}^{t_f} \Vert \nabla_u H(x(t),u(t),p(t)) \Vert^2 dt, \\
& \text{subject to}
& & \dot{p}(t)=-\nabla_x H(x(t),u(t),p(t)),\\
& & & p(t_f)=\mathbf{0}.
\label{hard_optimization}
\end{aligned}
\end{equation}
In the hard-constrained method\cite{molloy2019inverse}, a variable, $L(t)$, is considered such that $p(t)=L(t)c$. Thus, problem (\ref{hard_optimization}) is reformulated as:

\begin{equation}
\begin{aligned}
& \underset{c}{\text{minimize}}
& & c^TWc,\\
& \text{subject to}
& & \dot{L}(t)=-\nabla_x\phi^T(x(t),u(t))\\ & & &-\nabla_x f^T(x(t),u(t))L(t),  L(t_f)=\mathbf{0}.
\label{ioc_hard_optimization}
\end{aligned}
\end{equation}
Here, 
\begin{align}
W&=\int_{t_0}^{t_f} W_l^T(t)W_l(t) dt,
\label{W}
\end{align}
and
\begin{align}
W_l(t)&=\nabla_u\phi^T(x(t),u(t))+\nabla_u f^T(x(t),u(t))L(t).
\end{align}
The constraint equation on $L(t)$ in the optimization problem (\ref{ioc_hard_optimization}) is obtained from equation (\ref{cond1b}). The purpose of this section was to understand the hard-constrained minimum principle-based method. In the next two sections, this method will be extended to solve the inverse trajectory tracking optimal control problem for each follower pigeon in the flock.

\section*{IOC using single flight data} \label{single_trajectory}
In this section, the hard-constrained method will be extended for the inverse trajectory tracking optimal control problem for each follower homing pigeon. We now need to write  problem (\ref{original_optimal_problem}) in the structure of problem (\ref{general_optimal_problem}).  
\begin{equation}
\begin{aligned}
& \underset{U_i}{\text{minimize}}
& &  \int_{t_0}^{t_f} c_i^T\phi(X_i(t),U_i(t)) dt, \\
& \text{subject to}
& & \dot{X_i}(t)=f(X_i(t),U_i(t)),\\
& & & X_i(t_0)=X_{i_0},
\label{structured_optimal_problem}
\end{aligned}
\end{equation}
Thus, problem (\ref{structured_optimal_problem}) is essentially the same as problem (\ref{original_optimal_problem}) such that
\begin{equation}
    f(X_i(t),U_i(t))=\mathcal{A}X_i(t)+\mathcal{B}U_i(t),
\end{equation}
\begin{align}
c_i^T\phi(X_i(t),U_i(t))&=(X_{Ti}(t)-X_i(t))^TQ_i(X_{Ti}(t)-X_i(t))\notag\\&+U_i(t)^TR_iU_i(t),
\end{align}
\begin{equation}
    \phi(X_i(t),U_i(t))= \begin{bmatrix}(x_{Ti}(t)-x_i(t))^2 \\ (y_{T_i}(t)-y_i(t))^2 \\ (z_{T_i}(t)-z_i(t))^2) \\ (v_{x_{T_i}}(t)-v_{x_i}(t))^2 \\ (v_{y_{T_i}}(t)-v_{y_i}(t))^2 \\ (v_{z_{T_i}}(t)-v_{z_i}(t))^2\\u_{x_i}^2\\u_{y_i}^2\\u_{z_i}^2\end{bmatrix},
\end{equation}
and
\begin{equation}
    c_i=\begin{bmatrix}
        \hbox{vec}(Q_i)\\
        \hbox{vec}(R_i)
    \end{bmatrix}.
    \label{stack1}
\end{equation}
Here, $\hbox{vec}(\cdot)$ stacks the diagonal elements of a digonal matrix $(\cdot)$ into a vector. 

\begin{assu}
    It is assumed that the last element of $c_i$ is assumed to be known and equal to 1.
    \label{assu_ck}
\end{assu}

The IOC problem statement for single flight trajectory of each follow pigeon $i\in\{M,G,B,D,H,L,I,C,J\}$ will now be introduced.
\begin{prob}
    Given the system dynamics $\dot{X}_i(t)=f(X_i(t),U_i(t))$ for a follower pigeon $i\in\{M,G,B,D,H,L,I,C,J\}$, the basis function vector $\phi(X_i(t),U_i(t))$, and trajectory measurements, $X_i(t)$ and control signal $U_i(t)$, $\forall t_0\le t\le t_f$, which are measured from the primary optimal control problem (\ref{original_optimal_problem}), the objective of the inverse trajectory tracking optimal control problem is to find the weight vector $c_i$ for each follower pigeon $i\in\{M,G,B,D,H,L,I,C,J\}$.
    \label{goal_ioc2}
\end{prob}

The Hamiltonian function \cite{naidu2018optimal} for (\ref{structured_optimal_problem}) is:
\begin{align}
\mathcal{H}(X_i(t),U_i(t),p(t))&=c_i^T\phi(X_i(t),U_i(t))\notag\\&+p_i^T(t)f(X_i(t),U_i(t)),
\label{hamiltonian}
\end{align}
where $p_i(t)$ is the costate trajectory for the optimal control problem of the $i^{\hbox{th}}$ follower pigeon. If $(X_i(t),U_i(t))$ is optimal, that is, $(X_i(t),U_i(t))=(X^\star_i(t),U^\star_i(t))$, where $(X^\star_i(t),U^\star_i(t))$ is the optimal solution of problem (\ref{structured_optimal_problem}) for a given $c_i$, then there exists a costate trajectory, $p_i(t):\mathbb{R}\rightarrow\mathbb{R}^n$ (here, $p_i(t)=p^\star_i(t)$, where $p^\star_i(t)$ is the optimal costate trajectory) and it is necessary that \cite{johnson2013inverse}, \cite{naidu2018optimal}:
\begin{align}
\dot{p}^T_i(t)+c^T_i\nabla_{X_i}\phi\vert_{(X_i,U_i)}+p^T_i(t)\nabla_{X_i}f\vert_{(X_i,U_i)}=\mathbf{0}, \label{cond1}\\
c^T_i\nabla_{U_i}\phi\vert_{(X_i,U_i)}+p^T_i(t)\nabla_{U_i}f\vert_{(X_i,U_i)}=\mathbf{0}. \label{cond2}
\end{align}

\begin{figure*}
    \centering
\includegraphics[scale=0.7]{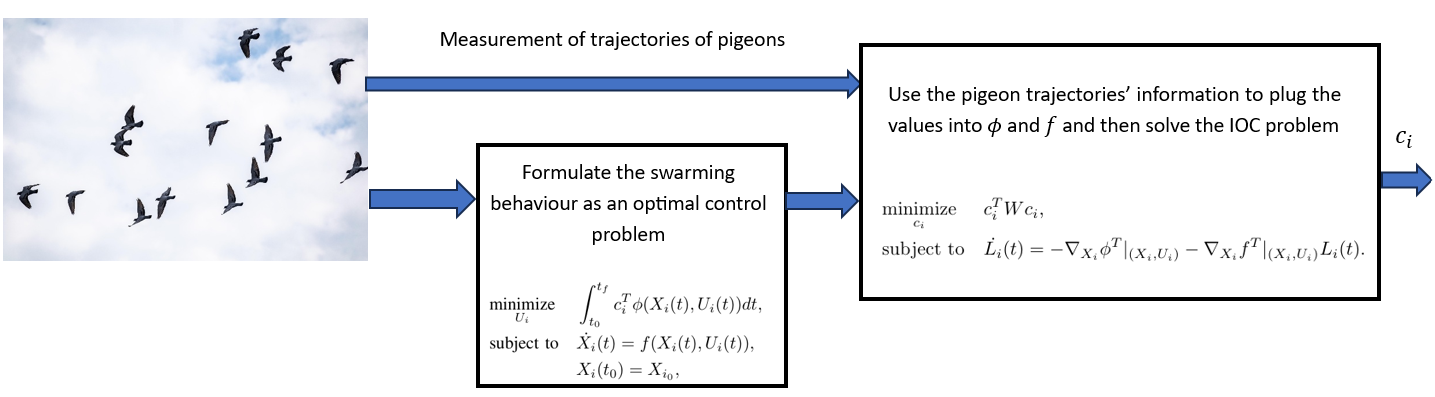}
    \caption{The swarming behaviour of the flock of pigeons is formulated as an optimal control problem. Measurements of trajectories of the pigeons are then used to solve the IOC problem to recover $c_i$}
    \label{method1}
    \end{figure*}

In minimum principle based methods, it is assumed that the measurements are only approximately optimal and not strictly optimal \cite{johnson2013inverse}. In the hard-constrained method \cite{molloy2019inverse}, (\ref{cond1}) is considered to be strictly satisfied while (\ref{cond2}) is considered to be satisfied only approximately. Hence, the left-hand side of (\ref{cond2}) is minimized assuming (\ref{cond1}) holds strictly. Thus, the optimization problem for this method becomes:
%\begin{equation}
%\begin{aligned}
%& \underset{c}{\text{minimize}}
%& &   \int_{t_0}^{t_f} \Vert \nabla_u H(t,x(t),u(t),p(t)) \Vert^2 dt, \\
%& \text{subject to}
%& & \dot{p}(t)=-\nabla_x H(t,x(t),u(t),p(t)),\\
%& & & p(t_f)=\mathbf{0}.
%\label{hard_optimization}
%\end{aligned}
%\end{equation}

\begin{equation}
\begin{aligned}
& \underset{c_i}{\text{minimize}}
& & c_i^TWc_i,\label{ioc_hard_optimization1}\\
& \text{subject to}
& & \dot{L_i}(t)=-\nabla_{X_i}\phi^T\vert_{(X_i,U_i)}\\& & &-\nabla_{X_i}f^T\vert_{(X_i,U_i)}L_i(t).
\end{aligned}
\end{equation}
where $L_i(t)$ is a variable such that $p_i(t)=L_i(t)c_i$,  $L_i(t_f)=\mathbf{0}$, and the constraint equation for $L(t)$ in  the optimization problem (\ref{ioc_hard_optimization1}) is obtained from equation (\ref{cond1}). The matrix $W$ in the optimization problem (\ref{ioc_hard_optimization1}) is given as:
\begin{align}
W&=\int_{t_0}^{t_f} W_1^T(t)W_1(t) dt, \label{W}\\
W_1(t)&=\nabla_{U_i}\phi^T\vert_{(X_i,U_i)}+\nabla_{U_i} f^T\vert_{(X_i,U_i)}L(t),
\end{align}
obtained from equation (\ref{cond2}).
The hard-constrained IOC method for single flight data for the follower pigeons can be understood from the block diagram in Fig. \ref{method1}. At first, the swarming behaviour of the flock of pigeons is formulated as an optimal control problem. The state and control trajectories of the pigeons are measured. The IOC problem is then solved using the real-time trajectories of pigeons being plugged into the functions $\phi(\cdot)$ and $f(\cdot)$ in the optimal control problem formulated.

\section*{IOC using multiple flight data for each follower pigeon} \label{multiple_trajectory}
In this section, we consider that multiple flight data are available for the flock of pigeons. The optimal trajectory tracking control problem for the $j^{\hbox{th}}$ flight number for the
$i^{\hbox{th}}$ follower pigeon is given as

\begin{equation}
\begin{aligned}
& \underset{U_{i_j}(t)}{\text{minimize}}
& & (X_{T_{i_j}}(t)-X_{i_j}(t))^TQ_i(X_{T_{i_j}}(t)-X_{i_j}(t))\\ & & &+U_{i_j}(t)^TR_iU_{i_j}(t)\\
& \text{subject to}
& & \dot{X_{i_j}}(t)=\mathcal{A}X_{i_j}(t)+\mathcal{B}U_{i_j}(t).
\label{original_optimal_problem2}
\end{aligned}
\end{equation}
\begin{assu}
    It is assumed that $Q_i\geq 0$ and $R_i>0$ are the same for every flight number $j$ for a particular follower pigeon $i\in\{M,G,B,D,H,L,I,C,J\}$
\end{assu}
We now need to structure problem (\ref{original_optimal_problem2}) in the form of problem (\ref{general_optimal_problem}).
\begin{equation}
\begin{aligned}
& \underset{U_{i_j}}{\text{minimize}}
& &  \int_{t_0}^{t_f} c_i^T\phi(X_{i_j}(t),U_{i_j}(t)) dt, \\
& \text{subject to}
& & \dot{X_i}(t)=f(X_{i_j}(t),U_{i_j}(t)),\\
& & & X_{i_j}(t_0)=X_{i_{j_0}},
\label{structured_optimal_problem2}
\end{aligned}
\end{equation}
Thus, problem (\ref{structured_optimal_problem2}) is essentially the same as problem (\ref{original_optimal_problem2}) such that
\begin{equation}
    f(X_{i_j}(t),U_{i_j}(t))=\mathcal{A}X_{i_j}(t)+\mathcal{B}U_{i_j}(t),
\end{equation}
\begin{figure*}[t!]
\begin{equation}
c_i^T\phi(X_{i_j}(t),U_{i_j}(t))=(X_{T_{i_j}}(t)-X_{i_j}(t))^TQ_i(X_{T_{i_j}}(t)-X_{i_j}(t))+U_{i_j}(t)^TR_iU_{i_j}(t),
\end{equation}
\hrulefill
\end{figure*}
\begin{equation}
    \phi(X_{i_j}(t),U_{i_j}(t))= \begin{bmatrix}(x_{T_{i_j}}(t)-x_{i_j}(t))^2 \\ (y_{T_{i_j}}(t)-y_{i_j}(t))^2 \\ (z_{T_{i_j}}(t)-z_{i_j}(t))^2) \\ (v_{x_{T_{i_j}}}(t)-v_{x_{i_j}}(t))^2 \\ (v_{y_{T_{i_j}}}(t)-v_{y_{i_j}}(t))^2 \\ (v_{z_{T_{i_j}}}(t)-v_{z_{i_j}}(t))^2\\u_{x_{i_j}}^2\\u_{y_{i_j}}^2\\u_{z_{i_j}}^2\end{bmatrix}.
\end{equation}
Here, $c_i$ is written as in (\ref{stack1}). The IOC problem statement for a flight number $j$ of each follow pigeon $i\in\{M,G,B,D,H,L,I,C,J\}$ will now be introduced.
\begin{prob}
    Given the system dynamics $\dot{X}_{i_j}(t)=f(X_{i_j}(t),U_{i_j}(t))$ for a follower pigeon $i\in\{M,G,B,D,H,L,I,C,J\}$ for a flight number $j$, the basis function vector $\phi(X_{i_j}(t),U_{i_j}(t))$, and trajectory measurements $X_{i_j}(t)$ and control signal $U_{i_j}(t)$, $\forall t_0\le t\le t_f$, which are measured from the primary optimal control problem (\ref{original_optimal_problem2}), the inverse trajectory tracking optimal control problem is to find the weight vector $c_i$ for each follower pigeon $i\in\{M,G,B,D,H,L,I,C,J\}$.
    \label{goal_ioc3}
\end{prob}
Assumption \ref{assu_ck} holds for the IOC problem using multiple flight trajectories. The Hamiltonian function \cite{naidu2018optimal} for (\ref{structured_optimal_problem2}) is:
\begin{align}
\mathcal{H}(X_{i_j}(t),U_{i_j}(t),p_{i_j}(t))&=c_i^T\phi(X_{i_j}(t),U_{i_j}(t))\notag\\&+p_{i_j}^T(t)f(X_{i_j}(t),U_{i_j}(t)),
\label{hamiltonian2}
\end{align}

Hence, we determine the cost function now based on multiple flight data for each leader-follower pair.  The necessary conditions of optimality for a flight number $j$ of a follower pigeon $i$ are given a \cite{johnson2013inverse}, \cite{naidu2018optimal}:
\begin{align}
\dot{p}_{i_j}^T(t)+c_i^T\nabla_{X_i}\phi\vert_{(X_{i_j},U_{i_j})}+p_{i_j}^T(t)\nabla_{X_{i_j}}f\vert_{(X_{i_j},U_{i_j})}=\mathbf{0}, \label{cond5}\\
c_i^T\nabla_{U_i}\phi\vert_{(X_{i_j},U_{i_j})}+p_{i_j}^T(t)\nabla_{U_i}f\vert_{(X_{i_j},U_{i_j})}=\mathbf{0}. \label{cond6}
\end{align}
By minimizing the left hand side of equation (\ref{cond6}) assuming equation (\ref{cond5}) holds strictly, the IOC problem for multiple flight trajectories become:
\begin{equation}
\begin{aligned}
& \underset{c_i}{\text{minimize}}
& & c_i^TWc_i,\label{ioc_hard_optimization2}\\
& \text{subject to}
& & \dot{L}_{i_j}(t)=-\nabla_{X_{i_j}}\phi^T\vert_{(X_{i_j},U_{i_j})}\\& & &-\nabla_{X_{i_j}}f^T\vert_{(X_{i_j},U_{i_j})}L_{i_j}(t).
\end{aligned}
\end{equation}
where $L_{i_j}(t)$ is a variable such that $p_{i_j}(t)=L_{i_j}(t)c_{i_j}$,  $L_j(t_f)=\mathbf{0}$,
\begin{align}
W&=\int_{t_0}^{t_f} W_2^T(t)W_2(t) dt, \label{Wnew}\\
W_2(t)&=\begin{bmatrix}
    W_{a_1}(t)\\W_{a_2}(t)\\ \vdots\\W_{a_j}(t)\\ \vdots\\W_{a_l}(t)
\end{bmatrix},\\
W_{a_j}(t)&=\nabla_{U_{i_j}}\phi^T\vert_{(X_{i_j},U_{i_j})}+\nabla_{U_i} f_{i_j}^T\vert_{(X_{i_j},U_{i_j})}L_{i_j}(t).
\end{align}
Here, $l$ are the total number of observations obtained from equation (\ref{cond6}).
Collecting multiple flight data that are inherently different for each flight can lead to a smaller condition number of $N_h^TWN_h$ as opposed to a larger condition number with a single flight data.

\begin{table*}
    \begin{center} 
    \begin{tabular}{|c|c|c|c|}
 \hline
\textbf{Flight No}  &  $\mathbf{t_f}$ & $c_1$, $c_2$, $c_3$, $c_4$, $c_5$, $c_6$, $c_7$, $c_8$ & $r_w$ \\
 \hline
 FF4  &  2451 s  & 0, 0, 0, 6.86, 5.09, 5.62, 59.06, 61.23 &  $2.33\times 10^{13}$         \\
 \hline
 FF5  & 2451 s   & 0, 0, 0, 2.84, 1.85, 3.90, 16.65, 15.35   &  $1.99\times 10^{12}$   \\
 \hline
 FF7  & 2451 s   & 0, 0, 0, 4.57, 3.65,  4.47, 49.97, 47.30    &   $2.36\times 10^{13}$   \\
  \hline
 FF9  &  2451 s  & 0, 0, 0, 2.24, 1.36, 3.27, 9.19, 9.19  & $3.55\times 10^{11}$    \\
 \hline
 FF4, FF5, FF7, FF9 & 2451 s & 0, 0, 0, 5.17, 3.66, 4.32, 39.11, 38.45 & $8.41 \times 10^{12}$ \\
 \hline
 \end{tabular}
\end{center}
\caption{\label{pigeon_M} Simulation results for pigeon M}
\end{table*}

 \begin{table*}
    \begin{center} 
    \begin{tabular}{|c|c|c|c|}
 \hline
\textbf{Flight No}  &  $\mathbf{t_f}$ & $c_1$, $c_2$, $c_3$, $c_4$, $c_5$, $c_6$, $c_7$, $c_8$ & $r_w$ \\
 \hline
 FF4  & 1000 s   &  0, 0, 0, 1.52, 0.52, 5.92, 12.52, 13.09   &   $1.16\times 10^{12}$  \\
 \hline
 FF5  &   1000 s  & 0, 0, 0, 0.96, 1.91, 3.83, 5.94, 5.30    &  $2.85\times 10^{11}$   \\
 \hline
 FF7  & 1000 s   & 0, 0, 0, 17.87, 10.90, 59.88, 279.39, 317.38    &   $7.35\times 10^{14}$  \\
 \hline
 FF9  &  1000 s  &  0, 0, 0, 0.81, 0.85, 0.85, 5.16, 4.68   & $1.29\times 10^{11}$    \\
 \hline
 FF4, FF5, FF7, FF9 & 1000 s  & 0, 0, 0, 5.32, 2.95,
 13.59, 39.50, 39.50 & $1.14 \times 10^{13}$\\
 \hline
 \end{tabular}

\end{center}
\caption{\label{pigeon_B} Simulation results for pigeon B}
\end{table*}

\begin{table*}
    \begin{center} 
    \begin{tabular}{|c|c|c|c|}
 \hline
\textbf{Flight No}  &  $\mathbf{t_f}$ & $c_1$, $c_2$, $c_3$, $c_4$, $c_5$, $c_6$, $c_7$, $c_8$ & $r_w$\\
 \hline
 FF4  &  2475.8 s  & 0, 0, 0, 3.93, 2.88, 9.72, 33.23, 34.42 &   $7.40\times 10^{12}$  \\
   \hline
 FF5  &  2475.8 s  & 0, 0, 0, 7.97, 7.63, 3.31, 40.56, 39.37    & $7.08\times 10^{12}$    \\
 \hline
 FF7  &  2475.8 s  & 0, 0, 0, 1.07, 0.94, 0.96, 9.42, 11.37    &  $4.04\times 10^{11}$   \\
 \hline
 FF9  & 2475.8 s  & 0, 0, 0, 1.62, 1.23, 1.08, 9.60, 9.56  &  $1.13\times 10^{12}$   \\
 \hline
 FF4, FF5, FF7, FF9 & 2475.8 s & 0, 0, 0, 3.35, 2.77, 2.77, 24.88, 26.23  & $3.64\times 10^{12}$\\
 \hline
 \end{tabular}

 \label{pigeon_C}

\end{center}
\caption{\label{pigeon_C} Simulation results for pigeon C}
\end{table*}

\section*{Uniqueness of the IOC solution}
The uniqueness of solution to the IOC problem obtained using the hard-constrained method will be reviewed here and shown for the swarm inverse optimal control problem. 
\begin{rem} \label{rem:solvability_hard}
  The solution of problem (\ref{ioc_hard_optimization1}) or (\ref{ioc_hard_optimization2}) exists and is unique if $W$ is unique and positive definite \cite[Theorem 3]{molloy2019inverse}. In optimal control problems, generally some of the parameters in the vector $c$ are fixed and known. Hence, this condition reduces to the principal minor of $W$ which excludes the columns for the known parameters \cite[Theorem 3]{molloy2019inverse}. An alternative compact formulation \cite{islam2025uniqueness} is: Given a vector $c \in \mathbb{R}^k$; a vector $\bar{c} \in \mathbb{R}^k$ can be found, populated only by non-zero elements at the corresponding indices for which the elements in $c$ are known. To ensure existence and uniqueness of the solution of problem (\ref{ioc_hard_optimization1}) or (\ref{ioc_hard_optimization2}), it is sufficient that $N_h^T W N_h$ is full rank for 
  \begin{equation}
  N_h=\ker\left(\text{diag}\left( \bar{c}  \right)\right). \label{equ:Nh}
  \end{equation}
\end{rem}

The theorem below is for uniqueness of the IOC solution.
\begin{thm}\label{Th:Theorem1}
Given the system dynamics $\dot{X}_i(t)=f(X_i(t),U_i(t))$ for a single flight data of a follower pigeon or $\dot{X}_{i_j}(t)=f(X_{i_j}(t),U_{i_j}(t))$  for multiple flight data of a pigeon, the basis function vector $\phi(X_i(t),U_i(t))$ or $\phi(X_{i_j}(t),U_{i_j}(t))$, and a specific system trajectory $X_i(t)$ or $X_{i_j}(t)$ and control signal $U_i(t)$ or $U_{i_j}(t)$, $\forall t_0\le t\le t_f$, measured from a follower pigeon $i\in\{M,G,B,D,H,L,I,C,J\}$ from a single flight data of an optimal trajectory tracking control problem (\ref{original_optimal_problem}) or flight number $j$ of multiple flights for the optimal problem (\ref{original_optimal_problem2}), the hard-constrained IOC problem (\ref{ioc_hard_optimization1}) or (\ref{ioc_hard_optimization2})  will have a unique solution, if $N_h^TWN_h$ is non-singular \cite[Theorem 1]{islam2025uniqueness}.

\normalfont \emph{Proof:} 
Similar to the proof of \cite[Theorem 3]{molloy2019inverse}, using \cite[Proposition 15.1]{gallier2011geometric} and \cite[Proposition 15.1]{gallier2011geometric} for quadratic optimization, it can be shown that both optimization problems (\ref{ioc_hard_optimization1}) and (\ref{ioc_hard_optimization2}) will have a unique solution if $N_h^TWN_h$ is non-singular or full rank. \hfill$\blacksquare$
%The proof is similar to \cite[Theorem 3]{molloy2019inverse}. Let $\bar{w}$ be the first column of $W$ with the last element removed and let the inverse of $W$ and $N_h^TWN_h$ be denoted as $W^{-1}$ and $\bar{W}^{-1}$. Using \cite[Proposition 15.1]{gallier2011geometric} for optimization of quadratic functions, the unique solution for optimization problems (\ref{ioc_hard_optimization1}) and (\ref{ioc_hard_optimization2}) is given as
%\begin{equation}
 %   \hat{c}_i=W^{-1}b,
  %  \label{proof}
%\end{equation}
%where $b$ is a vector obtained from the constraint equation in the optimization problems (\ref{ioc_hard_optimization1}) and (\ref{ioc_hard_optimization2}). Here, $\hat{c}_i$ is the unique solution of the optimization problem. Since the last element of $c$ is assumed to be known and chosen as 1, equation (\ref{proof}) becomes
%\begin{equation}
  %  \hat{c}_i=\begin{bmatrix}
   %     \bar{W}^{-1}b_{n-1}\\
   %     1
  %  \end{bmatrix}.
%\end{equation}

%$b_{n-1}$ is obtained by removing the last element of vector $b$. $\bar{W}^{-1}$ exists if $N_h^TWN_h$ is non-singular. This completes the proof.                
\label{thm:hard}
\end{thm}

\section*{Simulation Results}
For each follower pigeon, we use free-flight GPS data provided in the work by Nagy et al\cite{nagy2010hierarchical}. The GPS data of 10 homing pigeons are available with a sampling period of 0.2 s. High-resolution light weight GPS devices were employed \cite{nagy2010hierarchical}. For our analysis, we consider the following free-flight data from FF4, FF5, FF7 and FF9. We consider that each follower pigeon $i\in\{M,G,B,D,H,L,I,C,J\}$ is generating it's trajectories, viz, $x_i$, $y_i$, $z_i$, $v_{x_i}$, $v_{y_i}$, $v_{z_i}$, $a_{x_i}$, $a_{y_i}$ and $a_{z_i}$ by solving the optimal control problem (\ref{original_optimal_problem}) for single flight data. For each follower pigeon $i\in\{M,G,B,D,H,L,I,C,J\}$, the desired trajectory is the time-delayed trajectory of the corresponding leader in the leader-follower pair. The leader-follower pairs for a flock of 10 pigeons is shown in Fig. \ref{fig:hierarchy} and Table. \ref{leader_follower_pair}. For multiple flight data, it is considered that each flight $j$, the follower pigeon $i\in\{M,G,B,D,H,L,I,C,J\}$ is generating it's trajectories, viz, $x_{i_j}$, $y_{i_j}$, $z_{i_j}$, $v_{x_{i_j}}$, $v_{y_{i_j}}$, $v_{z_{i_j}}$, $a_{x_{i_j}}$, $a_{y_{i_j}}$ and $a_{z_{i_j}}$ by solving the optimal control problem (\ref{original_optimal_problem2}).

We will now present simulation results for the inverse optimal control problem for the follower pigeons. We consider each follower pigeon and present the recovered cost function using single flight trajectory and using multiple flight trajectory data as described in earlier sections. We decided to present the simulation results for pigeons M, B and C, because the results of all the pigeons are quite similar (The simulation results for the other pigeons are available and can be provided if requested). We denote $r_w$ as the condition number of $N_h^TWN_h$. We consider a long segment of the GPS data available. The condition number of $N_h^TWN_h$ for each of the cases is large. It does not become smaller for multiple trajectories case, because it appears that the flight data are similar for each flight in multiple flight trajectories for the follower pigeons. This does not lead to new information and the matrix $N_h^TWN_h$ is ill-conditioned. Nonetheless, the results presented here can be improved by using sufficiently different trajectories for each flight in future experiments which can help in leading to a well-conditioned matrix $N_h^TWN_h$. In such a case, the true weights will be found. 

Even though a unique solution has not been obtained in this analysis, we observe certain trends in the IOC solution. It appears that the weights on the trajectory tracking is negligible, whereas there is significant weightage on the control inputs. A more weightage on the control inputs means that the pigeons in the swarm might wish to save energy. There is also some weightage on the velocity tracking, which indicates that the optimal control problem for swarm behaviour might focus more on velocity tracking rather than trajectory tracking, with a significant weightage on the control inputs. 

This work however lays the foundation for formulating the swarm behaviour of a flock of birds as an optimal trajectory tracking control problem. The application of minimum principle based IOC methods for  swarm behaviour can help learn the philosopy behind how birds in a swarm generate their trajectories.

  \section*{Conclusion}
In this work, the leader-follower behaviour of a flock of pigeons was formulated as a trajectory tracking optimal control problem. The desired trajectory for each follower pigeon was considered to be the trajectory of the corresponding leader in the considered leader-follower pair of the swarm, delayed in time. Real-time GPS data of a flock of ten homing pigeons were analysed for different flight paths. The hard-constrained minimum principle based IOC method was extended for the swarm framework to analyse the trajectories of the follower pigeons. The purpose of using the IOC approach was to estimate the unknown weights of the cost function of the trajectory tracking optimal control problem for the follower pigeons in a flock. This work possibly lays the foundation for replicating swarm behaviour from a flock of birds into a swarm of multi UAV systems using inverse optimal control approaches.

%\begin{IEEEbiography}[{\includegraphics[width=1in,height=1.25in,clip,keepaspectratio]{Figures/photo.JPG}}]{Afreen Islam} is currently working as a Research Engineer at the University College Dublin in Ireland. She works with the Center for Space Research focusing on estimation and attitude control of satellites. She completed her MSc and PhD degrees from the University of Manchester in 2020 and 2024. Prior to this, she completed her master’s degree from the Indian Institute of Technology Guwahati in 2016. She worked in lecturing positions at two different universities in India before coming to Manchester. She was a recipient of the Neil Munro award for the best MSc dissertation at the University of Manchester. She was awarded a PhD scholarship from the University of Manchester and the British Council Scholarship for pursuing her MSc degree at Manchester. Her time at Manchester was more focused on developing the theoretical analysis for inverse optimal control, while currently, she is focused on applications of control systems for space systems.
%\end{IEEEbiography}

%\vspace{11pt}

%\bf{If you will not include a photo:}\vspace{-33pt}
%\begin{IEEEbiographynophoto}{John Doe}
%Use $\backslash${\tt{begin\{IEEEbiographynophoto\}}} and the %author name as the argument followed by the biography text.
%\end{IEEEbiographynophoto}

\vfill

\end{document}